# TRACKING THE 2011 STUDENT-LED MOVEMENT IN CHILE THROUGH SOCIAL MEDIA USE

Barahona, M., García, C., Gloor, P., & Parraguez, P.

P. Universidad Catolica de Chile
4860 Vicuna Mackenna Avenue
Santiago, Metro Region, 7820436, Chile
e-mail: cgarciah@uc.cl, pdparrag@uc.cl, pgloor@mit.edu, mibaraho@uc.cl

## ABSTRACT

Using social media archives of the 2011 Chilean student unrest and dynamic social network analysis, we study how leaders and participants use social media such as Twitter, and the Web to self-organize and communicate with each other, and thus generate one of the biggest "smart movements" in the history of Chile. In this paper we i) describe the basic network topology of the 2011 student-led social movement in Chile; ii) explore how the student leaders are connected to, and how are they seen by (a) political leaders, and (b) University authorities; iii) hypothesize about key success factors and risk variables for the Student Network Movement's organization process and sustainability over time. We contend that this social media enabled massive movement is yet another manifestation of the network era, which leverages agents' socio-technical networks, and thus accelerates how agents coordinate, mobilize resources and enact collective intelligence.

## CONTEXT: THE NETWORKED SOCIETY & STUDENT MOVEMENT

The world has been in a process of structural transformation for over two decades. This process is multidimensional, but it is associated with the emergence of a new technological paradigm, based on information and communication technologies, that emerged in the 1970s and are spreading around the world. Society shapes technology according to the needs, values, and interests of people who use the technology: "*Technology does not determine society: it expresses it. But society does not determine technological innovation: it uses it.*" (Castells, 1996/2000, p.15) Furthermore, information and communication technologies are particularly sensitive to the effects of social uses on technology. The history of the Internet provides ample evidence that the users, particularly the first few thousands, were, to a large extent, the producers of the technology. However, technology is a necessary, albeit not sufficient condition for the emergence of a new form of social organization based on networking, that is, the diffusion of networking in all realms of activity on the basis of digital communication networks.

Changes in the technologies, economic organization, and social practices of production in this networked environment have shaped new opportunities for how we create and exchange information, knowledge, and culture. These changes ''have increased the role of nonmarket and nonproprietary production, both by individuals alone and by cooperative efforts in a wide range of loosely or tightly woven collaborations'' (Benkler, 2005, p.5). These changes have increased the thresholds of available freedom for users, consumers and citizens, and thus hold great practical promise: as a dimension of individual freedom; as a platform for better democratic participation; and as a medium to foster a more critical and self-reflective culture. The student movement this paper focus on is an excellent test bed to precisely study the extent to which the above changes are verified or not in practice in emerging economies such as Chile.

In this paper we describe the preliminary topology of the Chilean networked student movement as it has been unfolding during 2011. We explore the network topology, emerging leaders, influencers and connectors by analyzing diachronic new media use and content in Twitter, and Google Insight archives. We contend the current networked smart movement in Chile is yet another example of collective intelligence enabled by social media in turbulent 2011.

Political opinions, and political behavior, are formed in the space of mediated communication. We live in a world of diversified and recombining messages in the communication space processed by both distributed and co-localized minds with increasingly autonomous sources of information. However, the domination of the media space over people's minds works through a fundamental mechanism: presence/absence of a message in the media space. (Castells, 2005) Everything or everyone that is absent from this space cannot reach the public mind, thus it becomes a non-entity. This binary mode of media politics has extraordinary consequences on the political process and on the institutions of society.

The network society, in the simplest terms, is a social structure based on networks operated by information and communication technologies based in microelectronics and digital computer networks that generate, process, and distribute information on the basis of the knowledge accumulated in the nodes of the networks. A network is a formal structure (Monge and Contractor, 2003). It is a system of interconnected nodes. Following the Internet design, societies, organizations and movements have evolved from bureaucratic/centralized to both decentralized and distributed networks. This evolving change towards de-centralization and democratization of decision-making has started to impact business, governments and society at large (Malone, 2004).

As the network society diffuses, and new communications technologies expand their networks, there is an explosion of horizontal networks of communication, quite independent from media business and governments, that allows the emergence of self-directed mass communications (Castells, 2005), smart mobs (Rheingold, 2006), participatory cultures (Jenkins, 2006) and collaborative innovation networks (Gloor, 2006). It is mass communication because it spreads through the Internet, so it potentially reaches whole countries and regions and eventually the whole planet. It is self-directed because it is often initiated by individuals or groups, bypassing the official media system. The explosion of user-generated content referred to as Web 2.0, including blogs, wikis, videoblogs, podcasts, social networking sites, streaming, and other forms of interactive, computer to computer communication sets up a new system of global, horizontal communication networks that, for the first time in history, allow people to communicate with each other without going through the channels set up by the institutions of society for socialized communication.

These new socio-technical conditions present both opportunities and challenges to the 'organizing process' (Weick, 1995; Malone, 2004) as well as to the democracy and society at large.

## CASE STUDY: METHODOLOGY, DATA & ACTORS

The emergence of online social networks opens up unprecedented opportunities to read the collective mind, discovering emergent trends while they are still being hatched by small groups of creative individuals. The Web has become a mirror of the real world, allowing researchers in predictive analytics to study and better understand why some new ideas change our lives, while others never make it from the drawing board of the innovator.

The embedding of the Internet into all aspects of society has led to (among other things) the widespread availability of a cacophony of fact, fiction and opinion composed by all elements of society: governments, news organizations and most importantly individuals. Taken together, these players represent a stream of the society consciousness, albeit a stream with many competing voices, agendas, and noise. From the range of sources it is clear that the opinions and activities that will move society and make news in the future are embedded in this data stream, tracking in real-time our digital traces (for example using Twitter, Facebook, etc.)

Through a combination of methods, this research attempts to map and analyze the Chilean student movement in perspective, considering both the macro and the micro network level. To focus the analysis and make it feasible, we have identified a validated set of students leading this movement who are acting as key public figures. They were identified both through their positions as officially elected representatives of the Chilean student federation (CONFECH) and by their public visibility as spoke-persons in the traditional and new media platforms.

In this way we define six clear leaders (with two very prominent figures: Camila Vallejo and Giorgio Jackson) and then considered their impact through social media (specially Twitter), mapping who they follow (as an indication of their influences) as well as their followers. We also analyzed their web presence to find clues about their organizational and communicational styles. Other Chilean educational leaders or prominent figures, but not just student leaders, were also analyzed in parallel as a way to compare their online behavior and style with the student leaders. The two most important in this context are Mario Waissbluth, senior lecturer and educational thought leader, founder of the citizen movement "Educación 2020" and Jaime Gajardo, president of the teachers union and member of the political commission of the communist party. They represent two extremes of a spectrum of senior figures actively involved with the movement and thus represent an interesting comparison point.

### Tools and Methods:
Through Twitter´s API and public resources such as Twiangulate (see http://www.twiangulate.com/) we have collected relational data and metadata in order to build tables and visualizations to uncover the underlying networks and facilitate its interpretation. The relations are constructed through "following" and "followers" (in Twitter terms) and the metadata includes fields like the subjects' self description, reported real name, number of followers and "following", location, etc. The total number of unique subjects (users) amount to more than 500 (different filters were applied at certain stages to facilitate the analysis).

With services such as Klout.com we were able to extract additional information in terms of online influence measures and compare the different subject's style. Since the number of Internet searches related to the student movement can be a powerful honest signal of the overall level of interest, we have also used Google Insights to identify trends and potential causalities.

For the analysis we are using Condor (Gloor & Zhao 2004), which allows us to easily collect Twitter, Blog, and Web data, and to calculate dynamic social network analysis metrics in combination with automatic content and sentiment analysis.

## **BRIEF HISTORY: FROM THE 2006 *PENGUIN REVOLUTION* TO THE 2011 STUDENT NETWORK MOVEMENT**

The "Penguin Revolution" was the first student citizen movement in Chile that was completely independent from existing political parties. It originated in the first months of the Michelle Bachelet Administration in May of 2006. Furthermore, the protagonists of this revolution were students, mainly from public high schools, who were between 15 and 18 years old. Street protests and national strikes conducted by them had big impact in the public opinion, mainly because of the appearance of students as relevant political actors, which was not seen since the dictatorship of Augusto Pinochet in the '80s. The main demand from these students was to improve the quality of public education and to assure a fair educational system. Finally, during June of 2006, President Bachelet, who had just gotten in power with a citizen-oriented government mantra, proceeded to address the short term demands of the student movement, in a backdrop of national strikes and deep critiques to the government. Moreover, this student network organization not only accomplished the ouster of the education minister, but also achieved the modification of the Organic Constitutional Law of Education (LOCE). One of the key strengths of this early student movement was the use of technological information dissemination tools, that is, both old and new media platforms by the different constituencies embracing this educational-oriented smart mob.

The new 2011 upsurge of popular unrest comes 18 months after a centre-right president, Sebastián Piñera, took office. Before, the centre-left Concertación had ruled for a relatively tranquil 20 years, overseeing a long and delicate transformation to democracy with a few exceptions such as the 2006 Penguin Revolution mentioned above. In 2011, thousands of high school and university students (some of them being 'grown-up penguins') marched through the capital's streets, as well as those of other major cities, demanding a radical overhaul of the education system. Education is not the only issue at stake, but it is certainly the main one.

The student marches have been far bigger than those organized by other protest groups. On several occasions, they have drawn 100,000 people to the streets. At the heart of the students' anger is a perception that Chile's education system is grossly unfair - that it gives rich students access to some of the best schooling in Latin America while dumping poor pupils in shabby, under-funded state schools. Out of the 65 countries that participated in the PISA tests, Chile ranked 64th in terms of segregation across social classes in its schools and colleges. This "educational apartheid" (Waissbluth, 2011) lies at the heart of the current unrest, and to a certain extent it has penetrated other domains.

## **SOCIAL MEDIA USE AND STUDENT MOVEMENTS IN CHILE**

Trying to establish a clear pattern between the new media platforms and the student "smart mobs", we found that in a relatively short period of time there was an important shift in the use of the predominant tools and means for coordination and communication. During the 2006 "penguin revolution", the key social media tools used to empower the student movement were both blogs and Fotologs. (Garcia, Urbina & Zavala, 2010). Although Internet platforms were used and Internet access was rapidly increasing, it was still not a mass phenomenon, the Web 2.0 paradigm was still evolving and mobile Internet access was a few years ahead. The contrast with the 2011 situation is noticeable, in that Internet access is no longer a privilege but almost considered a given. The mobile Internet is affordable reality for everybody, Web 2.0 platforms are the standard and they permeate everyday life (Silverstone, 2005; Ureta, 2008). This is illustrated in figure 1, showing the popularity of the primary sources used to coordinate the student movement and communicate its goals.

Figure 1 shows the 2010 tipping point in terms of media platforms. Twitter is now leaving behind the Fotolog platform (Chile was one of the most active countries using that service). Facebook is also a major means for communication, it surpassed Fotolog in mid 2008, representing today almost 30 times more interest (analyzed using the same methodology through Google search usage) than twitter, it is omitted in figure 1 to preserve the scale and facilitate the analysis.

Our analysis tells us that this technological paradigm shift has important consequences. For instance, communication has become much more mobile through interaction by affordable smartphones with Internet connection and direct and simple access to Facebook and Twitter.

We estimate that the horizontal and transparent flow of information under this emerging media system permeates the culture of the movement and fits its self-organizing, distributed nature particularly well. Another interesting discovery is illustrated in Figure 2: the relative interest in the word "blog" has fallen drastically in Chile since its peak in June 2008, to only 40% of its original value. As can be seen in figure 2, it coincides with the rise of Twitter, which now has more than double the interest than the word "blog". A possible explanation for this change in social media use could be the following: "The communications tools broadly adopted in the last decade are the first to fit human social networks well, and because they are easily modifiable, they can be made to fit better over time. Rather than limiting our communications to one-to-one and one-to-many tools, which have always been a bad feet to social life, we now have many-to-many tools that support and accelerate cooperation and action" (Shirky, 2009, p.55). In this sense, the Chilean media-enabled social movements evolved from a one-to-many paradigm (Fotolog), to a many-to-many one (Twitter).

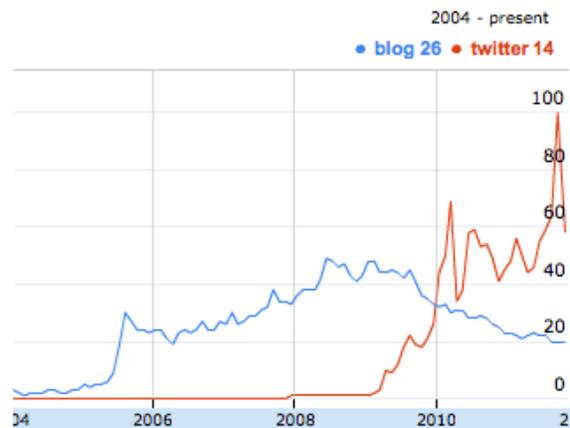

*Figure 1: Popularity of Google Web searches for terms "blog" and "twitter" in Chile*

According to our interviews with strategic informants—mainly with one of the leaders of the 2011 Student Movement, see appendix B—, this phenomenon has impacted the leaders of the current movement, who actively use Twitter for both organization and information diffusion. The twitter usage has been very effective, especially at the moment of coordinating the marches and other kinds of manifestations. The generalized use of twitter apps for mobiles allows the smart crowds to be always connected and up to date, being a fast-responsive, adaptive network. (Rheingold, 2002). The leaders have mostly abandoned the use of formal blogs. The new way offers higher capability for coordination plus an informational advantage when spreading news, but a diminished capability to create elaborated content, hindering the depth of the debate by limiting it to what can fit in 140 characters or links to videos or images.

This new use of media emphasizes the immediacy of communication as an evolving stream of information rather than as a cumulative stock of knowledge.

**Analysis: Twitter Data & Networks**

As a main component of our analysis we studied the twitter behavior of a set of key actors indentified as articulators and leaders of the current student movement. The selected actors are:

- Camila Vallejo: (born in Santiago, April 28, 1988) Geography graduate student and a leader of the student movement in Chile and the 2011 Chilean protests. She is a prominent member of the Communist Youth of Chile, and is the current president of the Student Federation of the University of Chile.

- Giorgio Jackson: (born in Santiago, February 6, 1987) is a student of Industrial Civil Engineering and leader of the student movement in Chile and the 2011 Chilean protest. In 2010 he was elected as president of the "FEUC" (Pontifical Catholic University of Chile Student Federation and acts as spokesman of the "CONFECH" (Chilean Student Confederation).
- Mario Waissbluth: Professor of the University of Chile, President and General Coordinator of "Educación2020". The movement has as objective to create a world-class education system for Chile before 2020.
- Freddy Fuentes: (born in Santiago, October 11, 1993) Spokesman of the "CONES" (High School Students Coordinator).
- José Ancalao: He is *werken*[1] of the "FEMAE" (Mapuche Student Federation).
- Alfredo Vielma: Spokesman at "ACES" (High School Coordinator Assembly).

Table 1 shows a series of indicators related to the activity in the network of these actors. It's important to notice that these actors (especially the young ones) have passed from practically being unknown to be Chilean-wide celebrities in a few months. Giorgio and Camila have "twitter similarity" with prominent businesspeople, media celebrities and politicians with long-term careers, such as the current Chilean president, Mr. Sebastián Piñera (measured for example in terms of their number of followers).

We aim to identify relevant people who are influencing the leaders of the student movement. In the "Twittersphere" these are the people who are followed by the actors we study. This motivation comes from observing the ratio of following/followers in twitter. As is shown in Table 1, the actors have an average ratio of 0,105. This can be interpreted as high selectivity by the key actors who are only following somebody if they are important as a source of information to them.

| Actor | Following | Followers | F/F | Tweets | Listed |
|---|---|---|---|---|---|
| Camila Vallejo | 108 | 324730 | 0,0003 | 405 | 1755 |
| Giorgio Jackson | 511 | 137643 | 0,0037 | 2826 | 604 |
| Mario Waissbluth | 581 | 49388 | 0,0118 | 30643 | 785 |
| Freddy Fuentes | 585 | 6752 | 0,0866 | 4478 | 53 |
| José Ancalao | 558 | 5875 | 0,0950 | 2254 | 58 |
| Alfredo Vielma | 86 | 198 | 0,4343 | 90 | 5 |
| Average | 404,8 | 87431,0 | 0,105 | 6782,6 | 543,3 |

*Table 1: Twitter indicators of six key actors*

In the first analysis, we obtained the "top 100" friends of the actors (see appendix A for the selection criteria we used). These results were however not satisfactory, mainly because we found mostly famous people such as TV and movie actors, show business people, etc.

We therefore changed the selection criteria of the twitter friends: instead of selecting the "top 100" we used an algorithm to select an "inner circle of friends". This algorithm is also described in appendix A. It removes the "famous people", giving preference to what might be "closer" friends. We suggest that these people are important to the student movement in two possible ways: their opinion is important to the leaders, or they are part of the movement's core team, acting as knowledge transfer agents.

---

[1] *Werken* could be translated as *Spokesman* from *Mapudungun*, the *Mapuche* language. The *Mapuches* were the main pre-Colombian culture present in Chile. According to the last Chilean census (2002), 4% of the population self-declared as *Mapuche*.

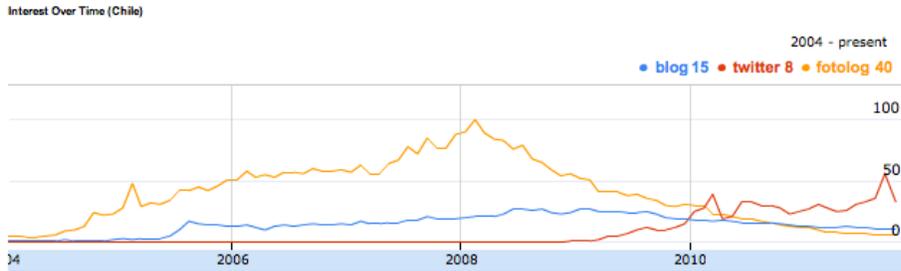

*Figure 2: Popularity of Google Web searches for social media technologies in Chile*

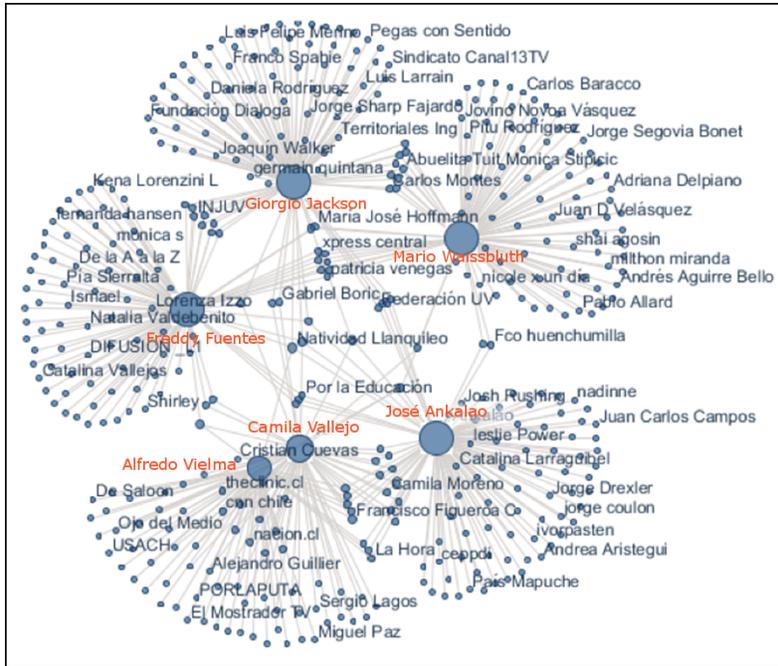

*Figure 3. Social network of six key actors constructed through their Twitter network*

In order to better understand the online behavior of some of the key leaders of the student movement, we also gathered information from KLOUT, an online tool to "measures influence based on the ability to drive action". This tool uses inputs from Twitter and other social networks to estimate aspects such as reach, amplification and how often top influencers respond to the content that is shared by the analyzed person.

Using the Klout score we obtained the following information:

| Leader | Score | Klout "style" | True Reach | Amplification | Network |
|---|---|---|---|---|---|
| Camila Vallejo | 75 | Thought Leader | 123K | 7 | 58 |
| Giorgio Jackson | 70 | Pundit | 59K | 7 | 55 |
| Jaime Gajardo | 62 | Broadcaster | 17K | 7 | 47 |
| Mario Waissbluth | 68 | Pundit | 37K | 8 | 53 |
| José Ankalao | 61 | Broadcaster | 14K | 12 | 50 |

*Table 2: Klout influence measures analysis. Explanation of each variable in Appendix C*

Through Table 2, it is possible to shed new light on important dimensions of online presence (mainly twitter) of the key leaders. It becomes clear that Camila Vallejo has a disproportionate amount of "true reach" (and also followers as stated in table 1). Nevertheless it is important to highlight that this is not directly transferred as amplification (how much the followers retweet or respond to the messages) or the relative influence of the followers (measured through the "network" variable).

The aggregated Klout score also shows that Mario Waissbluth, despite having a significantly smaller "true reach" and number of followers than Vallejo and Giorgio, has a similar aggregated score, due mainly to the amplification of his message that his followers provide and their respective influence. In this case it could be argued that due to the more senior and academic nature of Waissbluth's content, he can amplify his message across a relatively large number of influential individuals and elaborate arguments that generate conversations (instead of simple retweets).

**METADATA**

Every actor was classified in 9 categories (not mutually exclusive): Media, Student Leader, Student Movement, Political Representative, Social Leader, Government, Institutional Account, Artist and Academics [3]. The results in the sample are the following:

| Tag | Frequency |
|---|---|
| Media | 16 |
| Political Representative | 16 |
| Student Leader | 11 |
| Institutional Account | 9 |
| Social Leader | 5 |
| Government | 5 |
| Artists | 5 |
| Student Movement | 3 |
| Academics | 2 |

*Table 3: Frequency of every category using as sample the group of friends of more than one actor we study.*

In Figure 3, we see the six main actors selected in this case study: in the periphery we observe their inner circles and ego-networks. In the center of the graph, we observe their shared inner circles forming a star. These actors have two to five of the six main leaders in common.

**CONCLUSIONS**

The 2011 Chilean Student Movement case offers an excellent test bed to study the dynamics of network generation and diffusion. We encounter these self-organizing swarms and collaborative networks not only within the educational space, but also in the broader political space.

We found that Twitter is an excellent source to track the dynamics of both organization and diffusion of the student movement as described by the leaders and verified in our digital data set.

Our analysis of metadata obtained from Twitter profile information illustrates the relevance, interconnectivity and entanglement among the movement's leaders, political representatives, the media and key institutional agents. This collective student-based movement has, in fact, infected both the political and media systems by mobilizing resources, information and people across the public sphere, combining the digital and physical world.

We also observed two parallel tracks in the Chilean movement, which is clearly reflected in the two most important leaders. On the one hand, Giorgio Jackson defines the *analytic background* of the demands, focusing on the consistency of the arguments, the logic behind the debates and data driven discussions, using indicators and comparative data. His leadership style shows a sort of 'left brain' logic of thought and action, which is well characterized in his "Klout style" as a "pundit". On the other hand, Camila Vallejo has a rather *collective style*, very close to the youth and street-action. Her "Klout style" is "thought leader", which compared with Jackson, indicates a highly valued impact generated through less original content and online conversations. This duality of discursive styles allows them to reach a larger segment of the society than if only one of the styles would have been present.

**Further work**

In the near future, we expect to extend the interviews to other leaders of the movement, with in-depth questions aiming to understand the dynamics of the student movement, the diffusion across the student wide body as well as metrics of the temporal evolution of the movement.

In addition we will also include semantic analysis of Web news about the movement and its associated Wikipedia pages, to observe the evolution of the positions, demands and counter-offers in the negotiations between the student movement and the government. We will also do a sentiment analysis of social network contents, to get more information about the perception of the social movement in the Chilean population at large, including the political and higher education systems.

**APPENDIX**

**Appendix A: Definitions**
- Top 100 selections: This tool selects a maximum of 100 key followers of the actor. Key followers are calculated based on the influence and their own followers.
- Inner Network: Usually people with less followers but closer to the actor. These are people who might go undetected because they don't have a large number of followers on Twitter.

**Appendix B: Giorgio Jackson's answers:**

Using a scale from 1 to 7, where 1 is the lowest evaluation and 7 the highest, assign a score to the media with more impact in the **organization** (strategy, logistics, fundraising, etc.) of the students movement, which you have led:

| Media | Score |
|---|---|
| Facebook | 5 |
| Twitter | 6 |
| Email | 5 |
| Mobile | 4 |
| Blogs | 2 |
| Traditional media(newspapers, television, radio, etc) | 3 |

*Table 4: Usefulness of social media for organization ranked by Giorgo Jackson*

Using a scale from 1 to 7, where 1 is the lowest impact and 7 the highest, assign a score to the media with more impact in the **diffusion** (strategy, logistics, fundraising, etc.) of the students movement.

| Media | Score |
|---|---|
| Facebook | 6 |
| Twitter | 6 |
| Email | 4 |
| Mobile | 3 |
| Blogs | 4 |
| Traditional Media(newspapers, television, radio, etc) | 7 |

*Table 5: Usefulness of social media for information diffusion ranked by Giorgo Jackson*

**Appendix C: Klout variables**

- **True Reach:** How many people you influence
- **Amplification:** How much you influence them
- **Network Impact:** The influence of your network

**ACKNOWLEDGEMENTS**

The authors are grateful to Francisco Moya, Javier Urbina, and Patricia Hansen for their research input and suggestions. We also want to thank Giorgio Jackson, student leader, for his invaluable help and disposition to answer our questions.

**REFERENCES**

Benkler, Y. *The Wealth of Networks: How Social Production Transforms Markets and Freedom*. New Haven: Yale University Press, 2006.

Burt, Ronald. *Structural Holes: The Social Structure of Competition*. Cambridge, MA: Harvard University Press, 1992.

Castells, Manuel. *The Information Age: Economy, Society, and Culture* Volume 1: *The Network Society*. Oxford: Blackwell Publishers, 1996-2000.

García, C. Urbina, D. & Zavala, J. Social Media Meets Political Action: The 2006 Penguins Revolution in Chile. Working Paper PUC.

Gloor, P. Cooper, S. *Coolhunting* - Chasing Down The Next Big Thing? AMACOM, NY, 2007

Gloor, P. *Swarm Creativity, Competitive Advantage Through Collaborative Innovation Networks*. Oxford University Press, 2006

Gloor, P. Zhao, Y. TeCFlow: A Temporal Communication Flow Visualizer for Social Networks Analysis, ACM CSCW Workshop on Social Networks. ACM Conference, Chicago, Nov. 6. 2004.

Jenkins, H. Convergence *Culture: where old and new media collide*. New York University Press, 2006

Malone, Thomas. *THE FUTURE OF WORK: How the New Order of Business will Shape your Organization, your Management Style and your Life*. Boston: HBS Press, 2004.

Monge P. and Contractor, N. *Theories of Communication Networks*. OUP, 2004

Olson, M. *The Logic of Collective Action: Public Goods and the Theory of Groups*, Harvard University Press, Cambridge, Mass., 1971

Rheingold, H. *The Smart Mobs: The Next Social Revolution Transforming cultures and Communities in the Age of Instant Access*. Cambridge: Perseus Book Group, 2002.

Shirky, C. *Here Comes Everybody*, NY: Penguin Group, 2009.

Silverstone, R. *Media, technology and everyday life: from information to communication*. Ashgate England, 2005.

Ureta, S. *The Information Society,* Vol. 24, Iss. 2, 2008.

Waissbluth, M. *Se Acabó el Recreo: la desigualdad en la educación*, Ramdom House, Santiago, 2010.

Wassermann, S, Faust, K. *Social Network Analysis: Methods and Applications*, Cambridge University Press, 1994

Weick, K. *Sensemaking in Organizations*, NY: Sage Publications, 1995.

Zhang, X. Fuehres, H. Gloor, P. *Predicting Asset Value Through Twitter Buzz*. Altmann, J. Baumöl, U. Krämer, B. (eds) Proceedings 2nd. Symposium On Collective Intelligence Collin 2011, June 9-10, Seoul, Springer Advances in Intelligent and Soft Computing, vol. 112, 2011